\documentstyle{article}
\begin{document}
\LARGE
\begin{center}
\vspace*{0.3in} \bf The Cosmological Constant is Probably Zero, and
a Proof is Possibly Right

\vspace*{0.6in} \normalsize \large \rm
Zhong Chao Wu

Dept. of Physics

Zhejiang University of Technology

Hangzhou 310032, China

\vspace*{0.4in} \large \bf Abstract
\end{center}
\vspace*{.1in} \rm \normalsize \vspace*{0.1in}

Hawking proposed that the cosmological constant is probably zero in
quantum cosmology. Duff claimed that Hawking's proof is invalidated.
Using the right configuration for the wave function of the universe,
we provide a complete proof.

 \vspace*{0.8in}

PACS number(s): 04.60.-m; 98.80.Qc; 98.80.-k

Keywords: the cosmological constant, quantum cosmology, Euclidean
quantum gravity

\vspace*{0.5in}

\pagebreak

The largest discrepancy between theoretical calculations and
observations in the history of physics might be the value of the
cosmological constant. In order to resolve this, Hawking proposed in
quantum cosmology that ``the apparent cosmological constant is not
necessarily zero but that zero is by far the most probable value"
[1].

The contribution to the cosmological constant comes from the ground
states of all matter fields and ``the bare cosmological constant".
It is also known that a rank-3 antisymmetric tensor gauge field
$A_{\nu\rho\sigma}$ could contribute to the cosmological constant
too [2]. $A_{\nu\rho\sigma}$ arises naturally  in the $N=8$
supergravity in four dimensions. Hawking showed that when the total
cosmological constant becomes very small, the Euclidean action is
most negative, and the probability is therefore highest [1].

The relative creation probability of the universe is [3]
\begin{equation}
P \approx  \exp (-I),
\end{equation}
where $I$ is the Euclidean action of the seed instanton. The action
takes the form
\begin{equation}
I = -\int_M \left ( \frac{1}{16\pi}(R - 2\Lambda_0) - \frac{1}{48}
F^{\mu\nu\rho\sigma}F_{\mu\nu\rho\sigma} \right ),
\end{equation}
where the Planckian unit is used, $R$ is the scalar curvature,
$\Lambda_0$ represents the all contributions of ``the bare
cosmological constant" and the matter fields apart from
$A_{\nu\rho\sigma}$ [1], and $F$ is the field strength of
$A_{\nu\rho\sigma}$
\begin{equation}
F_{\mu\nu\rho\sigma}= \partial_{[\mu}A_{\nu\rho\sigma ]}.
\end{equation}
The gauge potential has the following gauge freedom
\begin{equation}
A_{\nu\rho\sigma} \longrightarrow  A_{\nu\rho\sigma} +
\partial_{[\nu}\lambda_{\rho \sigma ]}.
\end{equation}

Hawking argued that for the seed $S_4$ instanton the solution to the
gauge field equation
\begin{equation}
F^{\mu\nu\rho\sigma}_{\;\;\;\;\;\;\;\; ;\sigma} = 0
\end{equation}
should take the form
\begin{equation}
\sqrt{g}F^{\mu\nu\rho\sigma}= \kappa \epsilon^{\mu\nu\rho\sigma},
\end{equation}
where $\kappa$ is an arbitrary constant. In this model the $S_4$
instanton will evolve into the universe with the de Sitter spacetime
metric.

One can see from (2) that the $F^2$ term in the action behaves like
an effective cosmological constant
\begin{equation}
\Lambda_{eff}= 4\pi \kappa^2
\end{equation}
and the total cosmological constant is
\begin{equation}
\Lambda_{total} = \Lambda_0 + \Lambda_{eff}.
\end{equation}

The radius of $S_4$ is $(3/\Lambda_{total})^{1/2}$, and the action
is $- 3\pi/\Lambda_{total}$, here it is assumed that
$\Lambda_{total}$ is positive. The action is the negative of entropy
of the created de Sitter spacetime. From (1), it follows that the
most probable configuration will be those with very small values of
$\Lambda_{total}$, and nature will automatically select the right
value of $\kappa$ for this [1].

However, Duff pointed out that substituting a field configuration
into the action and varying it is not equivalent to substituting the
configuration into the field equations [4]. He explicitly showed
that for the configuration (6), the Einstein equation is
\begin{equation}
G_{\mu\nu} = 4\pi \kappa^2g_{\mu\nu} - \Lambda_0g_{\mu\nu}.
\end{equation}
This implies that from the field equation the total cosmological
constant must be $\Lambda_0 - 4\pi \kappa^2$, instead of $\Lambda_0
+ 4\pi \kappa^2$! Apparently, from observing the evolution of the
universe, the cosmological constant should take this value.

To recover the right effective cosmological constant in the action,
Aurelia, Nicolai and Townsend added a total divergence term to (2)
[2]
\begin{equation}
I_{div} =-\int_M dx^4\frac{1}{24}\kappa
\epsilon^{\mu\nu\rho\sigma}F_{\mu\nu\rho\sigma}.
\end{equation}
But in this case the value of $\kappa$, i.e, $\Lambda_{eff}$ is
fixed, in contradiction to the Hawking mechanism. Thus, this
prescription does not work for the cosmological constant issue.

Therefore, Duff claimed that ``this invalidates Hawking's proof that
the cosmological constant is probably zero" [4].

The motivation of this letter is to resolve this dilemma.

The probability expression (1) is derived from the wave function of
the universe [3], and the equator of the instanton and the other
fields at the equator are the configuration of the wave function.
For the action (2) one implicitly chooses 3-metric $h_{mn}$ of the
equator and $A_{\nu\rho\sigma}$ on it as the configuration. Indeed,
to derive the gauge field equation from the action (2), one has to
fix the value $A_{\nu\rho\sigma}$ at the boundary, i.e, the equator
in our case. In other words, if one simply uses the action (2) in
the no-boundary path integral, then the configuration of the wave
function of the universe should be $(h_{mn}, A_{\nu\rho\sigma})$.

In deriving the probability formula, one joins the south hemisphere
of the instanton and its time reversal, the north hemisphere, at the
equator. There is no way to get a regular $A_{\nu\rho\sigma}$ for
the whole $S_4$ in one piece. Instead, one can choose the gauge with
the regularity condition at the south hemisphere for
$A_{\nu\rho\sigma}$. The value at the north hemisphere is obtained
similarly via a sign change under the time reversal.  This results
in a discontinuity across the equator. Once the gauge is fixed, one
is not allowed to smooth it by a gauge transform (4). Therefore
$A_{\nu\rho\sigma}$ is not a right representation due to the
discontinuity across the equator, since deriving the probability (1)
from the two wave functions (for the north and south hemispheres)
one needs the same configuration from the two sides of the equator.
On the other hand, $F^{\mu\nu\rho\sigma}$ is a right representation
due to the continuity there.

One can always Fourier transform a wave function from one
representation to its conjugate in quantum theory in the Lorentzian
regime. In the Euclidean regime, at the $WKB$ level, this kind of
transform is equivalent to a Legendre transform of the instanton
action. The Legendre term is the summation of products of the
canonically conjugate variables at the boundary. For our model, the
term reads
\begin{equation}
I_{legendre}= -\int_{\Sigma_{S+N}} dS_\mu
\frac{1}{6}A_{\nu\rho\sigma}F^{\mu\nu\rho\sigma},
\end{equation}
where $\Sigma_{S+N}$ denotes the two equator boundaries for both the
south and north hemispheres.

Adding this term, the action (2) must be revised into
\begin{equation}
I = -\int_M \left ( \frac{1}{16\pi}(R - 2\Lambda_0) - \frac{1}{48}
F^{\mu\nu\rho\sigma}F_{\mu\nu\rho\sigma} \right
)-\int_{\Sigma_{S+N}} dS_\mu
\frac{1}{6}A_{\nu\rho\sigma}F^{\mu\nu\rho\sigma}.
\end{equation}
Apparently, varying the action (12) will result in the same field
equation (9) under the condition that $F^{\mu\nu\rho\sigma}$ is
given at the boundary. For our case, the boundary is $\Sigma_{S+N}$.

For the instanton, using the gauge field equation (5), one can
readily convert the Legendre term into the following form
\begin{equation}
I_{legendre}= -\int_M  \frac{1}{24}
F^{\mu\nu\rho\sigma}F_{\mu\nu\rho\sigma}
\end{equation}
and then the action (12) becomes
\begin{equation}
I = -\int_M \left ( \frac{1}{16\pi}(R - 2\Lambda_0) + \frac{1}{48}
F^{\mu\nu\rho\sigma}F_{\mu\nu\rho\sigma} \right )= -\int_M\left (
\frac{1}{16\pi}(R - 2\Lambda_0) + \frac{1}{2} \kappa^2 \right ).
\end{equation}

From the action (14) one can see that the $F^2$ term behaves as an
effective cosmological constant $-4\pi \kappa^2$ after substituting
the gauge field configuration of the instanton, which is the same as
what appears in the field equation (9) now. Therefore, as far as the
cosmological constant is concerned, Duff's dilemma has been
dispelled.

It is worth emphasizing that (12) and (14) are equivalent for the
instanton, or more accurately, the solution to the gauge field
equation. They are not equivalent for the more general case, since
we have used the gauge field equation (5) in deriving (14) from
(12).

After substituting the configuration, one is not allowed to consider
the gauge field as a variable again. However, considering the $F^2$
term in the action (14) as a constant for the given gauge field, one
can still vary the action with respect to the rest of the variables
and this results in the Einstein equation with a total cosmological
constant which is equivalent to (9), of course.  Everything is
consistent here.

In the above argument, it is implicitly assumed that
$\Lambda_{total}$, i.e, $R$ is positive. The Euclidean action is
obtained via the analytic continuation from the Lorentzian action.
There is a sign ambiguity in action (2) due to the continuation of
the factor $\sqrt{-g}$ from the Lorentzian action. The term
associated with $R$ in the Euclidean action must be negative, so
that the primordial fluctuation will take the ground state allowed
by the Heisenberg uncertainty principle [5]. By the same argument,
if $\Lambda_{total}$ or $R$ is negative, then the Euclidean action
should take the negative of the expressions (2)(12) and (14), and
the above argument remains intact. For both cases, the Euclidean
action can be written as $-3\pi/|\Lambda_{total}|$, and the
probability would exponentially increase no matter in which
direction the value of $\Lambda_{total}$ approaches zero. For the
case with negative $\Lambda_{total}$, the instanton is also $S_4$
but with a negative definite metric signature, and the created
universe is described by anti-de Sitter spacetime $AdS_4$.

One may wonder why the choice of representation in quantum cosmology
is so crucial. It is well known that one can equally use any
representation in quantum theory. For that case one is working in
the Lorentzian spacetime, while the quantum creation scenario occurs
in the Euclidean spacetime with imaginary time. In the Euclidean
regime, the no-boundary path integral with a wrong representation
does not make much sense.

Duff also said ``...cast doubt on similar attempts based on
maximizing the exponential of minus the Euclidean action" [4]. I
believe this statement can only be applied when one chooses the
wrong configuration or representation in the no-boundary path
integral.

One can interpret the path integral as the partition function in
gravitational thermodynamics [6], the right representation
corresponds to the microcanonical ensemble. This is very useful in
dealing with the problems of black hole with distinct surface
gravities or temperatures.

The representation of the wave function of the universe has been
previously discussed in the scenario of primordial black hole
creation [7] and spacetime dimensionality [8] in quantum cosmology.

\vspace*{0.3in} \rm

\bf Acknowledgement:

\vspace*{0.1in} \rm This work is supported by NSFC No.10703005.

\vspace*{0.3in} \rm

\bf References:

\vspace*{0.1in} \rm

1. S.W. Hawking, \it Phys. Lett. \rm \bf B\rm\underline{134}, 403
(1984).

2. M.J. Duff and P. van Nieuwenhuizen, \it Phys. Lett. \rm \bf
B\rm\underline{94}, 179 (1980); A. Aurilia, H. Nicolai and P.K.
Townsend, \it Nucl. Phys. \rm\bf B\rm\underline{176}, 509 (1980).

3. J.B. Hartle and S.W. Hawking, \it Phys. Rev. \rm \bf D\rm
\underline{28}, 2960 (1983).

4. M. Duff, \it Phys. Lett. \rm \bf B\rm \underline{226}, 36 (1989).

5. J.J. Halliwell and S.W. Hawking, \it Phys. Rev. \rm \bf D\rm
\underline{31}, 1777 (1985).

6. G.W. Gibbons and S.W. Hawking, \it Phys. Rev. \bf D\rm
\underline{15}, 2752 (1977).

7. S.W. Hawking and S.F. Ross,  \it Phys. Rev. \bf D\rm
\underline{52}, 5865 (1995); R.B. Mann and S.F. Ross, \it Phys. Rev.
\bf D\rm \underline{52}, 2254 (1995); Z.C. Wu, \it Int. J. Mod.
Phys. \rm \bf D\rm\underline{6}, 199 (1997), gr-qc/9801020; Z.C. Wu,
\it Phys. Lett. \rm \bf B\rm \underline{445}, 274 (1999),
gr-qc/9810012.

8. Z.C. Wu, \it Gene. Relativ. Grav. \rm\underline{34}, 1121 (2002),
hep-th/0105021; Z.C. Wu, \it Phys. Lett. \rm\bf B\rm\underline{585},
6 (2004), hep-th/0309178.

\end{document}